\documentclass[prd,twocolumn,preprintnumbers,nofootinbib,showpacs]{revtex4-1}
\usepackage{}
\usepackage{bm}
\usepackage{amsmath,amssymb}
\usepackage{rotating}
\usepackage{graphicx}
\usepackage{subfigure}

\newcommand{\beqn}{\begin{eqnarray}}
\newcommand{\eeqn}{\end{eqnarray}}
\newcommand{\eq}[1]{(\ref{#1})}

\usepackage{color}
\usepackage{float}

\begin{document}

\title{The effect of the chiral chemical potential on the chiral phase transition in the NJL model with different regularization schemes}

\author{Lang Yu$^{1,2}$}
\email{yulang@jlu.edu.cn}
\author{Hao Liu$^{2}$}
\email{haoliu@mail.ihep.ac.cn}
\author{Mei Huang$^{2,3}$}
\email{huangm@mail.ihep.ac.cn}
\affiliation{$^1$ Center of Theoretical Physics and College of Physics, Jilin University, Changchun, 130012, China}
\affiliation{$^2$ Institute of High Energy Physics, Chinese Academy of Sciences,
Beijing 100049, China}
\affiliation{$^3$ Theoretical Physics Center for Science Facilities,
Chinese Academy of Sciences,
Beijing 100049, China}
\date{\today}

\begin{abstract}
We study the chiral phase transition in the presence of the chiral chemical potential $\mu_5$ using the two-flavor
Nambu--Jona-Lasinio model. In particular, we analyze the reason why one can obtain two opposite behaviors of the chiral critical temperature
as a function of $\mu_5$ in the framework of different regularization schemes. We compare the modifications of the chiral condensate and the critical
temperature due to $\mu_5$ in different regularization schemes, analytically and numerically. Finally, we find that, for the conventional
hard-cutoff regularization scheme, the increasing dependence of the critical temperature on the chiral chemical potential is an artifact,
which is caused by the fact that it does not include complete contribution from the thermal fluctuations. When the thermal contribution
is fully taken into account, the chiral critical temperature should decrease with $\mu_5$.
\end{abstract}
\maketitle

\section{Introduction}

Quantum chromodynamics (QCD), the theory of strong interactions, is expected to explore the phase structure
and phase diagram of QCD. Such research on the QCD vacuum and matter is very important to get a deeper understanding of
the early Universe, the compact stars and the phenomena in relativistic heavy ion collision experiments (e.g. the quark-gluon plasma),
and so on. One of the most intriguing fundamental features of QCD is spontaneous breakdown of chiral symmetry. The
chiral symmetry breaking and restoration can be characteristically described in terms of the chiral condensate, which is the order
parameter of chiral symmetry breaking in the chiral limit (an approximate order parameter when the current quark mass is small).
As a consequence, it is crucial to make theoretical investigations on the behavior of the chiral condensate, in order to comprehend the
chiral phase transition and the nonperturbative properties of the strong interactions.

It is well known that QCD at zero and low temperatures has a nontrivial topological structure due to the existence of
certain gluon configuration, i.e. instantons~\cite{Belavin:1975fg,'tHooft:1976up,'tHooft:1976fv}, which can be assigned an integer-valued topological winding number.
And the infrared instanton structure is believed to explain chiral symmetry breaking of the QCD vacuum~\cite{Schafer:1996wv}. However, at present,
experimental evidence for the existence of such topological gluon configurations can only be found indirectly from the meson
spectrum~\cite{Witten:1979vv,Veneziano:1979ec}.

When at high temperatures, e.g. in the quark-gluon plasma, which is a phase of extremely hot QCD matter created in heavy ion collisions,
a copious production of another kind of topological gluon configurations (i.e. the QCD sphalerons) is expected~\cite{Manton:1983nd,Klinkhamer:1984di,Kuzmin:1985mm,Arnold:1987mh,Khlebnikov:1988sr,Arnold:1987zg}. The topologically
nontrivial sphaleron transitions can induce chirality imbalance, and thus lead to the breaking of the parity ($\mathcal {P}$) and charge-parity ($\mathcal {CP}$) symmetry in the hot plasma via the axial anomaly of QCD.
Since there is no direct $\mathcal {P}$ and $\mathcal {CP}$ violation in
QCD, chirality imbalance can only be produced locally and vanishes on average. It means that the probability to generate a local domain with a positive winding number $Q_W$ is the same as the probability to generate a local domain with a negative winding number $-Q_W$. And such local chirality imbalance can give rise to local $\mathcal {P}$- and $\mathcal {CP}$-odd
domains in the quark-gluon plasma phase of QCD. Unlike the instanton transitions, which are thermally suppressed~\cite{Manton:1983nd,Klinkhamer:1984di,Kuzmin:1985mm,Arnold:1987mh,Khlebnikov:1988sr,Arnold:1987zg}, the sphaleron transitions are not suppressed and so the gluon configurations with nonzero winding number can be produced with relatively high probability~\cite{McLerran:1990de,Moore:1997im,Moore:1999fs,Bodeker:1999gx,Moore:2000ara}. Therefore, the quark gluon plasma, created by the heavy ion collisions, provides the best platform to probe direct experimental evidence for the existence of nontrivial topological gluon configurations, as well as local
$\mathcal {P}$ and $\mathcal {CP}$ violation.

On the other hand, it has been suggested that strong magnetic fields can be generated in noncentral heavy ion collisions~\cite{Skokov:2009qp,Voronyuk:2011jd,Bzdak:2011yy,Deng:2012pc}. It was proposed that the interplay between the local chirality imbalance and the magnetic field will induce a current along the direction of the magnetic field, which is called chiral magnetic effect (CME)~\cite{Kharzeev:2007jp,Fukushima:2008xe,Fukushima:2010fe}. This effect leads to a separation of positive and negative electric charges with respect to the reaction plane of the heavy ion collisions. The recent observation of charge azimuthal correlations at RHIC and LHC~\cite{Abelev:2009ac,Abelev:2009ad,Abelev:2012pa} possibly resulted from the CME, which is a direct consequence of nontrivial topological gluon configurations in the strong magnetic field background. Moreover, Concerning the influence of the magnetic field on the QCD phase diagram, there have been many other interesting phenomena besides the CME, such as magnetic catalysis~\cite{Klevansky:1989vi,Klimenko:1990rh,Gusynin:1995nb}, inverse magnetic catalysis~\cite{Bali:2011qj,Bali:2012zg}, and vacuum superconductivity~\cite{Chernodub:2010qx,Chernodub:2011mc}. In particular, the local chirality imbalance induced by the nontrivial topological gluon configuration is also one of possible mechanisms trying to explain the inverse magnetic catalysis effect~\cite{Chao:2013qpa,Yu:2014sla,Yu:2014xoa}.

Since local chirality imbalance is expected to be produced in the quark gluon plasma, it is of great interest to investigate the effects of the chirality imbalance on the phase structure and phase transition of QCD.
In order to treat the induced chirality imbalance, an chiral chemical potential $\mu_5$ can be introduced, which creates a difference between
the number of right- and left-handed quarks. The chiral chemical potential
can be related to the $\theta$ angle of strong interactions as follows~\cite{Fukushima:2008xe}:
\beqn
 \mu_5=\frac{1}{2N_f}\frac{\partial \theta}{\partial t},
\eeqn
where $t$ is the time coordinate and $N_f$ is the number of the light flavors.

In this article, we will focus
on studying the effects of the chiral chemical potential $\mu_5$ on the chiral phase transition in the framework of the two-flavor Nambu--Jona-Lasinio (NJL) model. Actually, it has been studied by using some effective models of QCD~\cite{Fukushima:2010fe,Chernodub:2011fr,Gatto:2011wc,Cao:2015}, Dyson-Schwinger equations (DSEs)~\cite{Wang:2015tia,Xu:2015vna} and lattice QCD simulations~\cite{Braguta:2014ira} in recent years, but they do not reach an agreement on the modification of the chiral transition temperature by the chiral chemical potential $\mu_5$. The results of most effective models in Refs.~\cite{Fukushima:2010fe,Chernodub:2011fr,Gatto:2011wc} show that the critical temperature of the chiral phase transition decreases with $\mu_5$, whereas the results of the NJL model in Ref.\cite{Cao:2015}, DSEs in Ref.~\cite{Xu:2015vna}, and lattice QCD in Ref.~\cite{Braguta:2014ira} show a completely different dependence of the critical temperature on $\mu_5$, i.e. increasing with $\mu_5$. Since we have no idea about the details of computations in DSEs~\cite{Xu:2015vna} and lattice QCD~\cite{Braguta:2014ira}, we will not discuss them in this paper. It is very surprising that the results in Ref.~\cite{Fukushima:2010fe,Gatto:2011wc} are in contradiction to those in Ref.~\cite{Cao:2015}, although these two papers use the same NJL model. The only difference between them is the regularization scheme. Hence, at the same time, we will investigate how the regularization schemes, at least in the NJL model, influence the effects of the chiral chemical potential on the chiral condensate and the chiral critical temperature of QCD. This is the main scope of this article. Furthermore, in Refs.~\cite{Fukushima:2010fe,Chernodub:2011fr,Gatto:2011wc} the chiral phase transition becomes first order at some critical value of the chiral chemical potential, while in Ref.~\cite{Cao:2015} as well as \cite{Xu:2015vna,Braguta:2014ira}, there is no such a behavior and the order of the phase transition always stays the same. We find that the order of the transition also depends on the regularization scheme in the NJL model after doing numerical calculations.

The paper is organized as follows. In Sec.~\ref{model}, we give a
general description of the NJL model with introducing the chiral chemical
potential. In Sec.~\ref{results}, we perform a systematic analysis of the effects of the chiral chemical potential on the chiral condensate and the chiral critical temperature for different regularization schemes in the NJL model, analytically and numerically.  Finally, we present our conclusions in Sec.~\ref{conclusion}.

\section{Model}
\label{model}
The Lagrangian density of our model is given by
\beqn
 \mathcal{L} & = & \bar\psi \left( i\gamma_\mu \partial^{\mu}+\mu_5\gamma^0\gamma^5\right)\psi
  + G_S\left[\left(\bar\psi\psi\right)^2
  + \left(\bar\psi i \gamma^5 \bm\tau\psi\right)^2\right].\nonumber \\
\label{eq:L:basic}
\eeqn
In the above equation, $\psi$ corresponds to the quark field of two light flavors $u$ and $d$.
$G_S$ is the coupling constants with respect to the scalar and pseudoscalar channels. For simplicity, we will just work in the chiral limit throughout the paper, which gives no ambiguity for the definition of the chiral critical temperature. In addition, unlike Ref.~\cite{Fukushima:2010fe}, we work at zero magnetic field in this paper, so that we can avoid considering inverse magnetic catalysis, whose underlying physics is still under debate at present.

At the mean field level, the corresponding Lagrangian from Eq.~\eq{eq:L:basic} can be
given by the following formula:
\beqn
\mathcal{L} & = & -\frac{\sigma^2}{4G_S}
+\bar\psi\left( i\gamma_\mu  \partial^{\mu}-\sigma+\mu_5\gamma^0\gamma^5\right)\psi \;,
\label{eq:mean:field}
\eeqn
where $\sigma=-2G_S\langle \bar\psi\psi\rangle\,$ is the dynamical quark mass.

Thus, the thermodynamical potentials $\Omega^{S}$ with soft cutoff and $\Omega^{H}$ with hard cutoff are given by, respectively,
\beqn
\Omega^{S} &=&\frac{\sigma^2}{4G_S}-N_c N_f  \sum_{s=\pm}\int \frac{d^3p}{(2\pi)^3}f_\Lambda^2 (p) \, \omega_{s}(p)\,\nonumber \\&&-2N_c N_f\,T \sum_{s=\pm} \int\frac{d^3p}{(2\pi)^3}\ln\bigl( 1+\, e^{-\beta \omega_{s}} \bigr)\,\nonumber \\
&=&\frac{\sigma^2}{4G_S}-N_c N_f  \sum_{s=\pm}\int^{\infty}_{0}\frac{d p}{2\pi^2}f_\Lambda^2 (p) \, p^2 \omega_{s}\,\nonumber \\&&-2N_c N_f\,T \sum_{s=\pm} \int^{\infty}_{0}\frac{d p}{2\pi^2}\, p^2\ln\bigl( 1+\, e^{-\beta \omega_{s}} \bigr)\,,
\eeqn
and
\beqn
\Omega^{H} &=&\frac{\sigma^2}{4G_S}-N_c N_f   \sum_{s=\pm}\int\frac{d^3p}{(2\pi)^3} \, \omega_{s}(p)\,\nonumber \\&&-2N_c N_f\,T  \sum_{s=\pm}\int\frac{d^3p}{(2\pi)^3}\ln\bigl( 1+\, e^{-\beta \omega_{s}} \bigr)\,\nonumber \\
&=&\frac{\sigma^2}{4G_S}-N_c N_f  \sum_{s=\pm}\int^{\Lambda}_{0}\frac{d p}{2\pi^2} \, p^2\omega_{s}\,\nonumber \\&&-2N_c N_f\,T \sum_{s=\pm}\int^{\Lambda}_{0}\frac{d p}{2\pi^2}\, p^2\ln\bigl( 1+\, e^{-\beta \omega_{s}} \bigr)\,.
\eeqn
where $\omega_{s} \equiv  \omega_{s}(p)=\sqrt{\sigma^2 + \bigl[ |\bm p| + s\,\mu_5\,  \bigr]^2}$,
are the eigenvalues of the Dirac operator with spin factors $s=\pm 1$, and $\beta=1/T$. To avoid cutoff artifact, following Ref.~\cite{Fukushima:2010fe} we use a smooth regularization scheme, labeled as the soft cutoff, by introducing a form factor $f_\Lambda (p)$ to deal with the divergence in the zero-point energy:
\begin{equation}
 f_\Lambda(p) = \sqrt{\frac{\Lambda^{2N}}
  {\Lambda^{2N} + |\bm p|^{2N}}}\;,
\label{eq:f:p}
\end{equation}
where we take $N=5$ specifically. In the $N\rightarrow \infty$ limit, the above $f_\Lambda (p)$ (or $f_\Lambda (p)^2$) is reduced to the hard cutoff function $\theta(\Lambda-|\bm p|)$. Actually, we do not need to introduce a cutoff function for the thermal part of $\Omega$, since $\Omega_{th}$ is not divergent at all. However, in the conventional hard-cutoff regularization scheme, the hard cutoff function is always introduced for the thermal part of $\Omega$, which leads to a quite different behavior of the chiral condensate at finite temperature as well as the transition temperature $T_c$. This will be discussed later. Thus, we introduce a revised hard-cutoff regularization scheme $\Omega^{H'}$ by replacing the thermal part of $\Omega^{H}$ with that of $\Omega^{S}$:
\beqn
\Omega^{H'} &=&\frac{\sigma^2}{4G_S}-N_c N_f \int  \sum_{s=\pm}\frac{d^3p}{(2\pi)^3} \, \omega_{s}(p)\,\nonumber \\&&-2N_c N_f\,T  \sum_{s=\pm}\int\frac{d^3p}{(2\pi)^3}\ln\bigl( 1+\, e^{-\beta \omega_{s}} \bigr)\,\nonumber \\
&=&\frac{\sigma^2}{4G_S}-N_c N_f  \sum_{s=\pm}\int^{\Lambda}_{0}\frac{d p}{2\pi^2} \, p^2\omega_{s}\,\nonumber \\&&-2N_c N_f\,T \sum_{s=\pm}\int^{\infty}_{0}\frac{d p}{2\pi^2}\, p^2\ln\bigl( 1+\, e^{-\beta \omega_{s}} \bigr)\,.
\eeqn
It is natural to predict that the effects of the chiral chemical potential on the chiral condensate and the transition temperature obtained from $\Omega^{H'}$ should be consistent with those obtained from $\Omega^{S}$, Since we will find that $\Omega^{S}$ and $\Omega^{H/H'}$ give similar results in the QCD vacuum. These points will be discussed and verified in some details in the later section. Therefore, we will just need to show why $\Omega^{H'}$ leads to a result different from $\Omega^{H}$ about the influence of the chiral chemical potential on the transition temperature.

The thermodynamical potential $\Omega$ by using either soft cutoff or hard cutoff can be divided into two parts, the vacuum part and the thermal part:
\beqn
\Omega^{} &=& \Omega^{}_{vac}+\Omega^{}_{th}\,,
\eeqn
where
\beqn
\Omega^{S}_{vac}
&=&\frac{\sigma^2}{4G_S}-N_c N_f  \sum_{s=\pm}\int^{\infty}_{0}\frac{d p}{2\pi^2}f_\Lambda^2 (p) \, p^2 \omega_{s}\,\nonumber\\
\Omega^{S}_{th} &=& -2N_c N_f\,T \sum_{s=\pm} \int^{\infty}_{0}\frac{d p}{2\pi^2}\, p^2\ln\bigl( 1+\, e^{-\beta \omega_{s}} \bigr)\,,
\eeqn
and
\beqn
\label{eq:Omega:H}
\Omega^{H/H'}_{vac}
&=&\frac{\sigma^2}{4G_S}-N_c N_f  \sum_{s=\pm}\int^{\Lambda/\Lambda}_{0}\frac{d p}{2\pi^2} \, p^2 \omega_{s}\,\nonumber\\
\Omega^{H/H'}_{th} &=& -2N_c N_f\,T \sum_{s=\pm} \int^{\Lambda/\infty}_{0}\frac{d p}{2\pi^2}\, p^2\ln\bigl( 1+\, e^{-\beta \omega_{s}} \bigr)\,,\nonumber\\
\eeqn

Now, $\sigma$ can be determined self-consistently by
solving the saddle point equation (the gap equation)
\beqn
\frac{\partial\Omega}{\partial\sigma}=0.
\label{eq:Omega:min}
\eeqn

\section{chiral phase transition with chiral chemical potential}
\label{results}

In this section we compare the results on the QCD chiral phase transition at zero and finite chiral chemical potential by using three different regularization schemes, i.e.  $\Omega^{S}$, $\Omega^{H}$ and $\Omega^{H'}$.
In this way we analyze the reasons why $\Omega^{H'}$ (or $\Omega^{S}$) and $\Omega^{H}$ exhibit similar behaviors on the chiral condensate and the transition temperature at zero and low temperatures, but totally different ones at high temperatures.

In addition, our model parameter set is
\beqn
\Lambda&=&626.76 \text{MeV},\,~~~~~~~G_S\Lambda^2=2.02,\,~~~~\text{for}\,~ \Omega^{S};\\
\Lambda&=&633.27 \text{MeV},\,~~~~~~~G_S\Lambda^2=2.19,\,~~~~\text{for}\,~ \Omega^{H}\,~ \text{and}\,~\Omega^{H'}\nonumber\\.
\eeqn
These parameters correspond to $f_{\pi}=92.3$ MeV and the constituent quark mass $M=325$ MeV.

\subsection{Analysis and Results at $\mu_5=0$}

\begin{figure}

   \centering

   \includegraphics[width=0.49\textwidth]{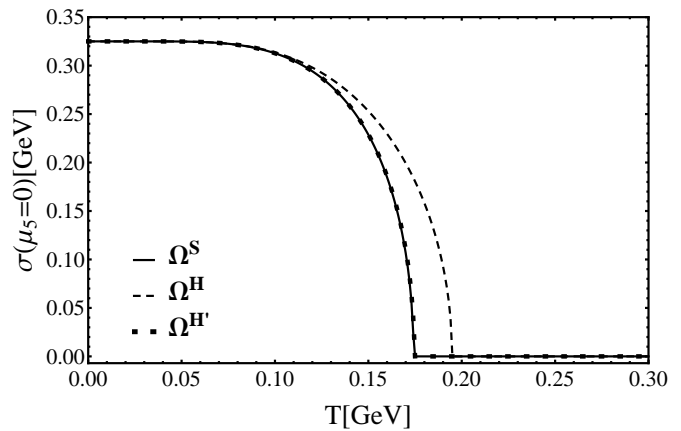}
\caption{ The dynamical quark mass $\sigma$ as a function $T$ at $\mu_5=0$ for $\Omega^{S}$, $\Omega^{H}$ and $\Omega^{H'}$ . }
  \label{fig:sigma:mu50}

\end{figure}

When at $\mu_5=0$, with the help of numerical simulations, we compute the behavior of the dynamical quark mass $\sigma$ as a function $T$ for $\Omega^{S}$, $\Omega^{H}$ and $\Omega^{H'}$, which is shown in Fig.~\ref{fig:sigma:mu50}. And numerically we find the transition temperature $T_c$ for these three regularization schemes: $T^{S}_c(\mu_5=0)=174.3~\text{MeV}$ for
$\Omega^{S}$, $T^{H}_c(\mu_5=0)=194.9~\text{MeV}$ for
$\Omega^{H}$, and $T^{H'}_c(\mu_5=0)=174.6~\text{MeV}$ for
$\Omega^{H'}$. Apparently, the results by $\Omega^{S}$ are consistent with those by $\Omega^{H'}$, but the values of $T_c$ and $\sigma$ near $T_c$ by $\Omega^{H}$ are larger than those by $\Omega^{S}$ and $\Omega^{H'}$.

As we know, at $\mu_5=0$, $\Omega^{H}_{vac}$ and $\Omega^{H'}_{vac}$ have the same expressions, and we obtain that
\beqn
\Omega^{H/H'}_{vac}(\mu_5=0)
&=&\frac{\sigma^2}{4G_S}-\frac{N_c N_f}{2\pi^2}\times\frac{1}{4}\Big[2\Lambda^3\sqrt{\Lambda^2+\sigma^2}\nonumber\\&&+\Lambda
\sigma^2\sqrt{\Lambda^2+\sigma^2}+\sigma^4\ln(\frac{\sigma}{\Lambda
+\sqrt{\Lambda^2+\sigma^2}})\Big]\nonumber \\
\eeqn
When $\sigma/\Lambda \rightarrow 0$, we can get
\beqn
\label{eq::vacH:mu50}
 \Omega^{H/H'}_{vac}(\mu_5=0)
 &\sim& \frac{\sigma^2}{4G_S}-\frac{3}{2\pi^2}(\Lambda^4+\Lambda^2\sigma^2
 +\frac{1}{8}\sigma^4\nonumber\\&&+\frac{1}{2}\sigma^4\ln\frac{\sigma}{2\Lambda})\nonumber \\
 &=&(\frac{1}{4G_S}-\frac{3\Lambda^2}{2\pi^2})\sigma^2-\frac{3\Lambda^4}{2\pi^2}
 -\frac{3}{16\pi^2}\sigma^4\nonumber\\&&+\frac{3}{4\pi^2}\sigma^4\ln\frac{2\Lambda}{\sigma}.
\eeqn
The coefficient of $\sigma^2$ term, i.e. $(\frac{1}{4G_S}-\frac{3\Lambda^2}{2\pi^2})$, must be negative, since $\sigma=0$ is a local maximum at QCD vacuum and we can easily verify it by using the fitting parameter value of $G_S$ ( $\frac{1}{4G_S}-\frac{3\Lambda^2}{2\pi^2}=-0.0378\Lambda^2$). And when $\sigma/\Lambda \rightarrow \infty$, we have
\beqn
\Omega^{H/H'}_{vac}(\mu_5=0)\sim \frac{\sigma^2}{4G_S}-\frac{2\Lambda^3}{\pi^2}\sigma.
\eeqn
It is easy to find that the value of $\sigma$ at the local minimum position, i.e. $\sigma(T=0)=\sigma_{min}=\frac{4G_S\Lambda^3}{\pi^2}$, is an approximate solution of the gap equation at $T=0$. And, of course, it is easy to check that this a very rough approximate result, but it reveals the existence of a local minimum solution qualitatively and that $\sigma_{min}\propto G_S$.

When at finite temperature ($\mu_5=0$ here), $\Omega^{H}_{th}$ and $\Omega^{H'}_{th}$ will give quite a different result on the transition temperature because of their different cutoff schemes. And we find that $T_c$ obtained by $\Omega^{H}_{th}$ is higher than that obtained by $\Omega^{H'}_{th}$ because of the different upper limits of the integrals. By using the high temperature expansion~\cite{Dolan:1973qd,Haber:1981fg,Haber:1981tr} of $\Omega^{H'}_{th}(\mu_5=0)$ (when $\sigma\ll T$), we have
\beqn
\Omega^{H'}_{th}(\mu_5=0)
&=&
-2N_c N_f\Bigg\{\frac{7\pi^2T^4}{360}-\frac{T^2\sigma^2}{24}+\frac{\sigma^4}{16\pi^2}\Big[\ln(\frac{\pi T}{\sigma})\nonumber\\&&-\gamma_E+\frac{3}{4}\Big]+\mathcal {O}(\frac{\sigma^6}{T^2})\Bigg\},\nonumber \\
\eeqn
where $\gamma_E=0.5772...$ is Euler's constant. When $T\rightarrow T_c$, $\sigma\ll T$ and $\sigma\ll \Lambda$, and thus we have
\beqn
\Omega^{H'}(\mu_5=0) &=&\Omega^{H'}_{vac}(\mu_5=0)+\Omega^{H'}_{th}(\mu_5=0)\nonumber \\
&\approx&\Bigg[(\frac{1}{4G_S}-\frac{3\Lambda^2}{2\pi^2})\sigma^2-\frac{3\Lambda^4}{2\pi^2} -\frac{3}{16\pi^2}\sigma^4\nonumber\\&&+\frac{3}{4\pi^2}\sigma^4\ln\frac{2\Lambda}{\sigma}\Bigg]
-12\Bigg\{\frac{7\pi^2T^4}{360}-\frac{T^2\sigma^2}{24}\nonumber \\
&&+\frac{\sigma^4}{16\pi^2}\Big[\ln(\frac{\pi T}{\sigma})-\gamma_E+\frac{3}{4}\Big]\Bigg\},\nonumber \\
&=&(\frac{1}{4G_S}-\frac{3\Lambda^2}{2\pi^2}+\frac{T^2}{2})\sigma^2-\frac{3\Lambda^4}{2\pi^2}
-\frac{7\pi^2T^4}{30}\nonumber\\&&-\frac{3\sigma^4}{4\pi^2}\Big[\ln(\frac{\pi T}{2\Lambda})-\gamma_E+1\Big].
\eeqn
When the coefficient of $\sigma^2$ term changes the sign from negative to positive as $T$ increases, $\sigma=0$ point becomes a local minimum and the corresponding temperature could be assumed to be the transition temperature $T_c$ approximately. In this way we obtain $T_c(H')\approx\sqrt{\frac{3\Lambda^2}{\pi^2}- \frac{1}{2G_S}}$ for $\Omega^{H'}(\mu_5=0)$. By using fitting parameters, it is easy to get $T_c(H')=174.6\, \text{MeV}$ for $\mu_5=0$, which is very close to the fully numerical result.

If we use the conventional hard cutoff for the thermal part of potential, i.e. $\Omega^{H}_{th}(\mu_5=0)$, a higher value of $T_c$ will be found.
\beqn
\Omega^{H}(\mu_5=0) &=&\Omega^{H}_{vac}(\mu_5=0)+\Omega^{H}_{th}(\mu_5=0)\nonumber \\
&\approx&\Bigg[(\frac{1}{4G_S}-\frac{3\Lambda^2}{2\pi^2})\sigma^2-\frac{3\Lambda^4}{2\pi^2}
-\frac{3}{16\pi^2}\sigma^4\nonumber\\&&+\frac{3}{4\pi^2}\sigma^4\ln\frac{2\Lambda}{\sigma}\Bigg]
-12\chi^2(\Lambda)\Bigg\{\frac{7\pi^2T^4}{360}-\frac{T^2\sigma^2}{24}\nonumber \\
&&+\frac{\sigma^4}{16\pi^2}\Big[\ln(\frac{\pi T}{\sigma})-\gamma_E+\frac{3}{4}\Big]\Bigg\},\nonumber \\
&=&(\frac{1}{4G_S}-\frac{3\Lambda^2}{2\pi^2}+\chi^2(\Lambda)\frac{T^2}{2})\sigma^2-\frac{3\Lambda^4}{2\pi^2}
\nonumber\\&&-\chi^2(\Lambda)\frac{7\pi^2T^4}{30}-\chi^2(\Lambda)\frac{3\sigma^4}{4\pi^2}\Big[\ln(\frac{\pi T}{2\Lambda})-\gamma_E+1\Big].\nonumber \\
\label{eq:H:highT}
\eeqn
where $\chi^2(\Lambda)=\Omega^{H}_{th}(\mu_5=0)/\Omega^{H'}_{th}(\mu_5=0)$. Since the integrands of $\Omega^{H}_{th}(\mu_5=0)$ and $\Omega^{H'}_{th}(\mu_5=0)$ are the same and positive, $\chi^2(\Lambda)< 1$. Hence, for $\Omega^{H}(\mu_5=0)$, we get $T_c(H)\approx\frac{1}{\chi(\Lambda)}\sqrt{(\frac{3\Lambda^2}{\pi^2}- \frac{1}{2G_S})}>T_c(H')$.
 Therefore, $T_c(H)>T_c(H')$ at $\mu_5=0$ is due to the fact that the conventional cutoff scheme neglects the contribution of the integral from $\Lambda$ to $\infty$ in the thermal part of the potential. It will be seen in the next subsection that, when at finite $\mu_5$, this is also the reason why $T_c$ increases with $\mu_5$ in $\Omega^{H}$ (when $\mu_5 <0.4$ GeV actually) , while $T_c$ decreases with $\mu_5$ in $\Omega^{H'}$.

\subsection{Analysis and Results at $\mu_5 \neq 0$}

\begin{figure}

   \centering

   \includegraphics[width=0.49\textwidth]{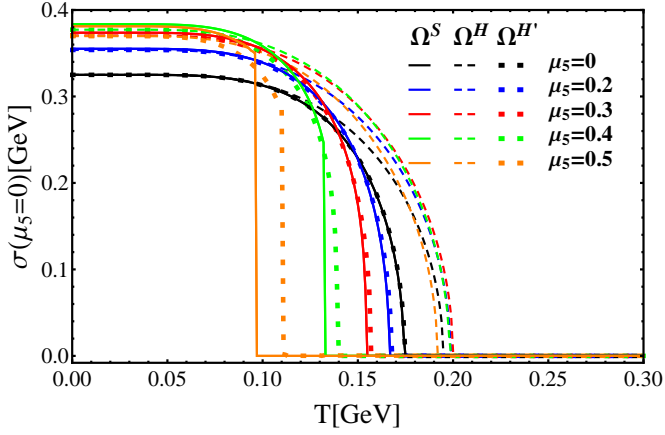}
\caption{ The dynamical quark mass $\sigma$ as a function $T$ at $\mu_5=0,~0.2,~0.3,~0.4~\text{and}~0.5$ for $\Omega^{S}$, $\Omega^{H}$ and $\Omega^{H'}$ . }
  \label{fig:sigma:mu5}

\end{figure}

When $\mu_5$ is introduced, in Fig.~\ref{fig:sigma:mu5}, we firstly display our numerical results about $\sigma$ as a function of $T$ at $\mu_5=0.2,~0.3,~0.4,~\text{and}~0.5$ GeV for three different regularization schemes, comparing with the case at $\mu_5=0$. From Fig.~\ref{fig:sigma:mu5}, we observe that, if $\mu_5$ is not too large (less than $0.4$ GeV at least), the chiral condensate increases with $\mu_5$ for all regularization schemes when at zero and low temperatures, which agrees with the result obtained in Refs.~\cite{Fukushima:2010fe,Chernodub:2011fr,Gatto:2011wc,Cao:2015,
Wang:2015tia,Xu:2015vna,Braguta:2014ira}. But it is no longer valid for $\Omega^{S}$ and $\Omega^{H'}$ when the temperature approaches $T_c$ and this result is just what was found in Refs.~\cite{Fukushima:2010fe,Chernodub:2011fr,Gatto:2011wc}. And if $\mu_5$ is large enough, we notice that, the chiral condensate at zero temperture begins to decrease with $\mu_5$ and this result has never been found in any previous literature, which will be discussed later. On the other hand, for $\Omega^{S}$ and $\Omega^{H'}$, the effect of $\mu_5$ is to lower the chiral transition temperature, which is exactly the same result in Ref.~\cite{Fukushima:2010fe,Gatto:2011wc}. As for $\Omega^{H}$, there exists a critical $\mu_5^*$ between 300 MeV and 400 MeV: $\mu_5$ is to raise the critical temperature of the chiral phase transition when $\mu_5 <\mu_5^* $ GeV, which agrees with the result in Ref.~\cite{Cao:2015}, while $\mu_5$ turns to decrease the transition temperature when $\mu_5 >\mu_5^*$ GeV, in contrast to the result in Ref.~\cite{Cao:2015}.

In addition, we note that,
there exists a critical $\mu_5^c$ in $\Omega^{S}$ between 300 MeV and 400 MeV, above which the chiral phase transition becomes first-order from second-order, which is just as same as the result in Ref.~\cite{Fukushima:2010fe}. And also we find $\Omega^{H'}$ have similar behavior for the order of the chiral phase transition, but its critical $\mu_5^c$ is between 400 and 500 MeV. In contrast, the chiral phase transition of $\Omega^{H}$ is always second-order, which is found in Ref.~\cite{Cao:2015}. It indicates that whether there is a change from second-order to first-order above a critical $\mu_5^c$ might be due to the choice of the special regularization scheme in the NJL model. And we will not discuss it detailedly in this paper.

Based on above numerical results, now we begin to explore the puzzle that why different regularization schemes in the NJL model give different effects of $\mu_5$ on the chiral phase transition temperature, especially between $\Omega^{H}(\mu_5)$ and $\Omega^{H'}(\mu_5)$ (since $\Omega^{H'}(\mu_5)$ and $\Omega^{S}$ have similar results). At $\mu_5\neq0$, $\Omega^{H}(\mu_5)$ and $\Omega^{H'}(\mu_5)$ can be expressed in the following way:
\beqn
\Omega^{H/H'}(\mu_5) &=& \Omega^{H/H'}(\mu_5=0)+\delta\Omega^{H/H'}(\mu_5),
\eeqn
where $\delta\Omega^{H/H'}(\mu_5)=\Omega^{H/H'}(\mu_5)-\Omega^{H/H'}(\mu_5=0)$,
and it can be decomposed into $\delta\Omega^{H/H'}_{vac}(\mu_5)+\delta\Omega^{H/H'}_{th}(\mu_5)$.
$\delta\Omega^{H/H'}_{vac}(\mu_5)$ and $\delta\Omega^{H/H'}_{th}(\mu_5)$ contain the effects of the chiral chemical potential $\mu_5$ on the chiral phase transition for the vacuum and the thermal parts, respectively. And we will analyze them analytically and numerically in this paper. In the following, analytical derivations will be given in two feasible scenarios: one is small $\mu_5$ expansion, which is valid at $\mu_5\ll \sigma$, the other is high-temperature with nonzero $\mu_5$, which is valid at $\sigma \ll T$ and $\mu_5 \ll T$. Note that $\mu_5$ cannot be too large in both scenarios, and if $\mu_5$ is large enough, the influence of the chiral chemical potential can only be computed and analyzed numerically, which will be shown and discussed in the following also.

\subsubsection{Small $\mu_5$ expansion}

When $\mu_5$ is small enough ($\mu_5\ll \sigma$), we can expand the thermodynamical potential in powers of $\mu_5$ up to the second order:
\beqn
\Omega &=& \Omega^{(0)}+\Omega^{(1)}+\Omega^{(2)}+\cdot\cdot\cdot
\eeqn
And it is easy to find that $\Omega^{(1)}=0$, and the nonzero components $\Omega^{(0)}$ and $\Omega^{(2)}$ in hard cutoff and soft cutoff are given as follows, respectively,
\beqn
\Omega^{H/H'(0)}&=& \Omega^{H/H'(0)}_{vac}+\Omega^{H/H'(0)}_{th}\,,
\eeqn
\beqn
\Omega^{H/H'(2)}&=& \Omega^{H/H'(2)}_{vac}+\Omega^{H/H'(2)}_{th}
\eeqn
with
\beqn
\Omega^{H/H'(0)}_{vac} &=& \frac{\sigma^2}{4G_S}-N_c N_f  \sum_{s=\pm}\int^{\Lambda/\Lambda}_{0}\frac{d p}{2\pi^2} \, p^2 \omega^{(0)}_{s}\,,\\
\Omega^{H/H'(0)}_{th} &=& -2N_c N_f\,T \sum_{s=\pm} \int^{\Lambda/\infty}_{0}\frac{d p}{2\pi^2}\, p^2\ln\bigl( 1+\, e^{-\beta \omega^{(0)}_{s}} \bigr)\,,\nonumber\\ \\
\Omega^{H/H'(2)}_{vac}&=&
\label{fig:H2:vac}
-\mu_5^2\,\sigma^2\Bigg[N_c N_f\sum_{s=\pm}\int^{\Lambda/\Lambda}_{0}\frac{d p}{2\pi^2} \, \frac{p^2 }{2(p^2+\sigma^2)^{3/2}}\Bigg]\,,\nonumber\\ \\
\Omega^{H/H'(2)}_{th} &=& \mu_5^2\,N_c N_f\Bigg\{\sum_{s=\pm}
\label{fig:H2:the}
 \int^{\Lambda/\infty}_{0}\frac{d p}{2\pi^2}\,\bigg[\frac{\sigma^2}{(1+e^{\beta \omega^{(0)}_{s}})}\nonumber\\&&\times \frac{p^2}{(p^2+\sigma^2)^{3/2}}- \frac{e^{\beta \omega^{(0)}_{s}}}{(1+e^{\beta \omega^{(0)}_{s}})^2}\frac{p^4}{T\,(p^2+\sigma^2)}\bigg]\Bigg\}\nonumber\\
\eeqn
and
\beqn
\Omega^{S(0)}&=& \Omega^{S(0)}_{vac}+\Omega^{S(0)}_{th}\,,\nonumber\\
\Omega^{S(2)}&=& \Omega^{S(2)}_{vac}+\Omega^{S(2)}_{th}
\eeqn
with
\beqn
\Omega^{S(0)}_{vac} &=& \frac{\sigma^2}{4G_S}-N_c N_f  \sum_{s=\pm}\int^{\infty}_{0}\frac{d p}{2\pi^2} \,f_\Lambda^2 (p)  p^2 \omega^{(0)}_{s}\,,\\
\Omega^{S(0)}_{th} &=& -2N_c N_f\,T \sum_{s=\pm} \int^{\infty}_{0}\frac{d p}{2\pi^2}\, p^2\ln\bigl( 1+\, e^{-\beta \omega^{(0)}_{s}} \bigr)\,,\nonumber\\ \\
\Omega^{S(2)}_{vac}&=&
-\mu_5^2\,\sigma^2\Bigg[N_c N_f\sum_{s=\pm}\int^{\infty}_{0}\frac{d p}{2\pi^2} \, f_\Lambda^2 (p) \frac{p^2 }{2(p^2+\sigma^2)^{3/2}}\Bigg]\,,\nonumber\\ \\
\Omega^{S(2)}_{th} &=& \mu_5^2\,N_c N_f\Bigg\{\sum_{s=\pm}
 \int^{\infty}_{0}\frac{d p}{2\pi^2}\,\bigg[\frac{\sigma^2}{(1+e^{\beta \omega^{(0)}_{s}})}\nonumber\\&&\times \frac{p^2}{(p^2+\sigma^2)^{3/2}}- \frac{e^{\beta \omega^{(0)}_{s}}}{(1+e^{\beta \omega^{(0)}_{s}})^2}\frac{p^4}{T\,(p^2+\sigma^2)}\bigg]\Bigg\}\nonumber\\
\eeqn
where $\omega^{(0)}_{s} =  \sqrt{\sigma^2 +  |\bm p| ^2}$.

At QCD vacuum ($T=0$), we can obtain from the above equations that the chiral chemical potential $\mu_5$ enhances the coupling constant $G_S$
effectively:
\beqn
\frac{1}{4G^{eff}_S(\mu_5)}&=&\frac{1}{4G_S}-\mu_5^2\,\Bigg[N_c N_f\sum_{s=\pm}\int^{\Lambda}_{0}\frac{d p}{2\pi^2} \, \nonumber\\&&\times\frac{p^2 }{2(p^2+\sigma^2)^{3/2}}\Bigg]\,\, \text{for}\, \Omega^{H}~ \text{and}~\Omega^{H'},\\
\frac{1}{4G^{eff}_S(\mu_5)}&=&\frac{1}{4G_S}-\mu_5^2\,\Bigg[N_c N_f\sum_{s=\pm}\int^{\infty}_{0}\frac{d p}{2\pi^2} \,\nonumber\\&&\times f_\Lambda^2 (p)  \frac{p^2 }{2(p^2+\sigma^2)^{3/2}}\Bigg]\,\, \text{for}\, \Omega^{S},
\eeqn
where $G^{eff}_S(\mu_5)$ is the effective coupling constant for the scalar channel with nonzero $\mu_5$, which is larger than $G_S$ obviously and increases with $\mu_5$. Hence, the chiral chemical potential $\mu_5$ contributes to enhance the chiral condensate at zero temperature. From another point of view, when $\sigma\rightarrow 0$, we can find that $\Omega^{H/H'(2)}_{vac}$ helps to increase the magnitude of the coefficient of $\sigma^2$ term for $\Omega^{H/H'}_{vac}$, since the coefficient of $\sigma^2$ term for $\Omega^{H/H'(0)}_{vac}$ by Eq.~(\ref{eq::vacH:mu50}), i.e. $(\frac{1}{4G_S}-\frac{3\Lambda^2}{2\pi^2})$, is negative. And $\Omega^{S}_{vac}$ can give the same result, since the coefficient of $\sigma^2$ term for $\Omega^{S(0)}_{vac}$ must be negative also.

\begin{figure}

   \centering

   \includegraphics[width=0.49\textwidth]{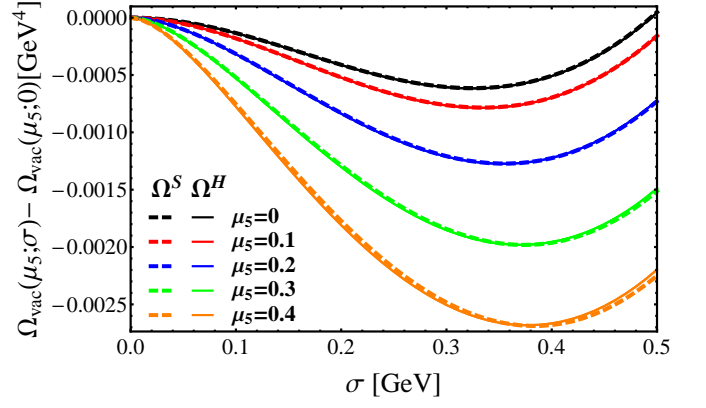}
\caption{ The potential $\Omega_{vac}(\mu_5;\sigma)-\Omega_{vac}(\mu_5;0)$ as a function of $\sigma$  at $\mu_5=0,\,0.1\,,0.2\,,0.3\,\text{and}\,0.4$ GeV. }
  \label{fig:1}

 \end{figure}

\begin{figure}
   \centering

   \includegraphics[width=0.49\textwidth]{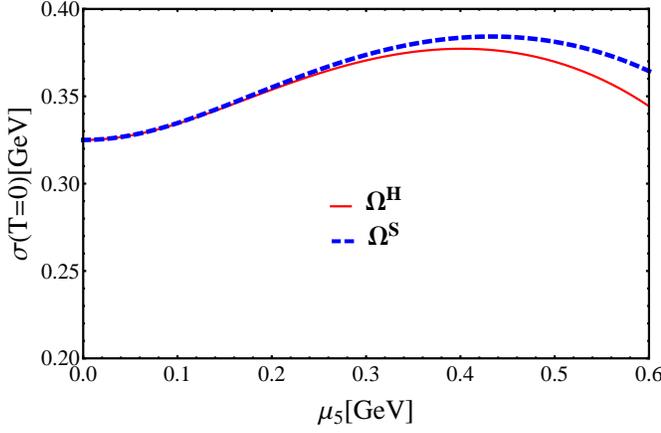}
\caption{The chiral condensate at zero temperature as a function of $\mu_5$ for $\Omega_S$ and $\Omega_H$(or $\Omega_{H'}$).}
 \label{fig:2}

\end{figure}

Although the above result is obtained at $\mu_5\ll \sigma$, we find that it remains valid even when $\mu_5$ is big enough from numerical results. In Fig.~\ref{fig:1}, we display the potential $\Omega_{vac}(\mu_5;\sigma)-\Omega_{vac}(\mu_5;0)$ as a function of $\sigma$ at $\mu_5=0,\,0.1\,,0.2\,,0.3\,\text{and}\,0.4$ GeV for $\Omega^S$ and $\Omega^{H/H'}$ by doing numerical calculations. Obviously, the greater $\mu_5$ values, the faster the curves decrease from the origin. It means that $\delta\Omega_{vac}(\mu_5)$ helps to increase the magnitude of the coefficient of $\sigma^2$ term for $\Omega$ when $\sigma \rightarrow 0$, which is in agreement with what we get from small $\mu_5$ expansion. In another word, the chiral condensate in the QCD vacuum will increase with $\mu_5$ accordingly, and it is also shown by Fig.~\ref{fig:1} that the minimum of the curve shifts to a larger value of $\sigma$ as $\mu_5$ increases. Thus, $\mu_5 < 0.4$ GeV, increasing $\mu_5$ results in slight enhancement of the chiral condensate at zero temperature, which is verified by the numerical computations and shown in Fig.~\ref{fig:2} as well. However, from Fig.~\ref{fig:2}, we find another interesting phenomenon that  there exists a threshold value of $\mu_5$ ($\mu^{thr}_5$ is about between $0.4$ and $0.5$ GeV), above which the quark condensate starts to decrease with $\mu_5$.

In addition, one can find that the results of the effect of a finite $\mu_5$ on the chiral condensate in the QCD vacuum, by using either $\Omega^{S}_{vac}$ or $\Omega^{H/H'}_{vac}$, agree with each other qualitatively and quantitatively (the difference between them is very small). It is expected that $\Omega^{S}$ and $\Omega^{H'}$ will give consistent results at both zero and finite temperature, since $\Omega^{S}_{th}$ and $\Omega^{H'}_{th}$ have the same expressions. Therefore, in this paper we just need to compare $\Omega^{H'}$ with $\Omega^{H}$ at finite temperature, as mentioned in the previous section, in order to find the reasons for their different effects of $\mu_5$ on chiral transition temperature.

However, at finite temperature, we cannot easily get such a simple result as at zero temperature, because we have to consider the difference between $\delta\Omega^{H}_{th}(\mu_5)$ and $\delta\Omega^{H'}_{th}(\mu_5)$. Of course, when at low temperatures, it is natural to expect that $\delta\Omega_{vac}(\mu_5)$ will give dominant contribution for either $\Omega^H$ or $\Omega^{H'}$, and thus helps to enhance the chiral condensate as $\mu_5$ increases in both regularization schemes, like at zero temperature. When at high temperatures ($T \gtrsim T_c$), it becomes complicated. In the small $\mu_5$ limit, by using $\delta\Omega^H_{th}(\mu_5)=\Omega^{H(2)}_{th}+\cdot\cdot\cdot$ and Eq.~\ref{fig:H2:the}, one can find that the first term of $\Omega^{H(2)}_{th}$ in Eq.~\ref{fig:H2:the} makes an opposite contribution to the coefficient of $\sigma^2$ term as a result of its positive sign, as compared with $\Omega^{H(2)}_{vac}$ in Eq.~\ref{fig:H2:vac}. Hence, $\Omega^{H(2)}_{th}$ will help to decrease the chiral condensate. Moreover, since $\frac{1}{(1+e^{\beta \omega^{(0)}_{s}})} < \frac{1}{2}$ at any temperature, the contribution of $\Omega^{H(2)}_{th}$ to the magnitude of the coefficient of $\sigma^2$ term is always less than that of $\Omega^{H(2)}_{vac}$. As a consequence, $\Omega^{H(2)}$ helps to increase the chiral condensate at any temperatures, and thus the chiral transition temperature $T_c$ will increase with $\mu_5$ for $\Omega^H$ correspondingly in the small $\mu_5$ limit.

\begin{figure}
 \centerline{\includegraphics[scale=0.49]{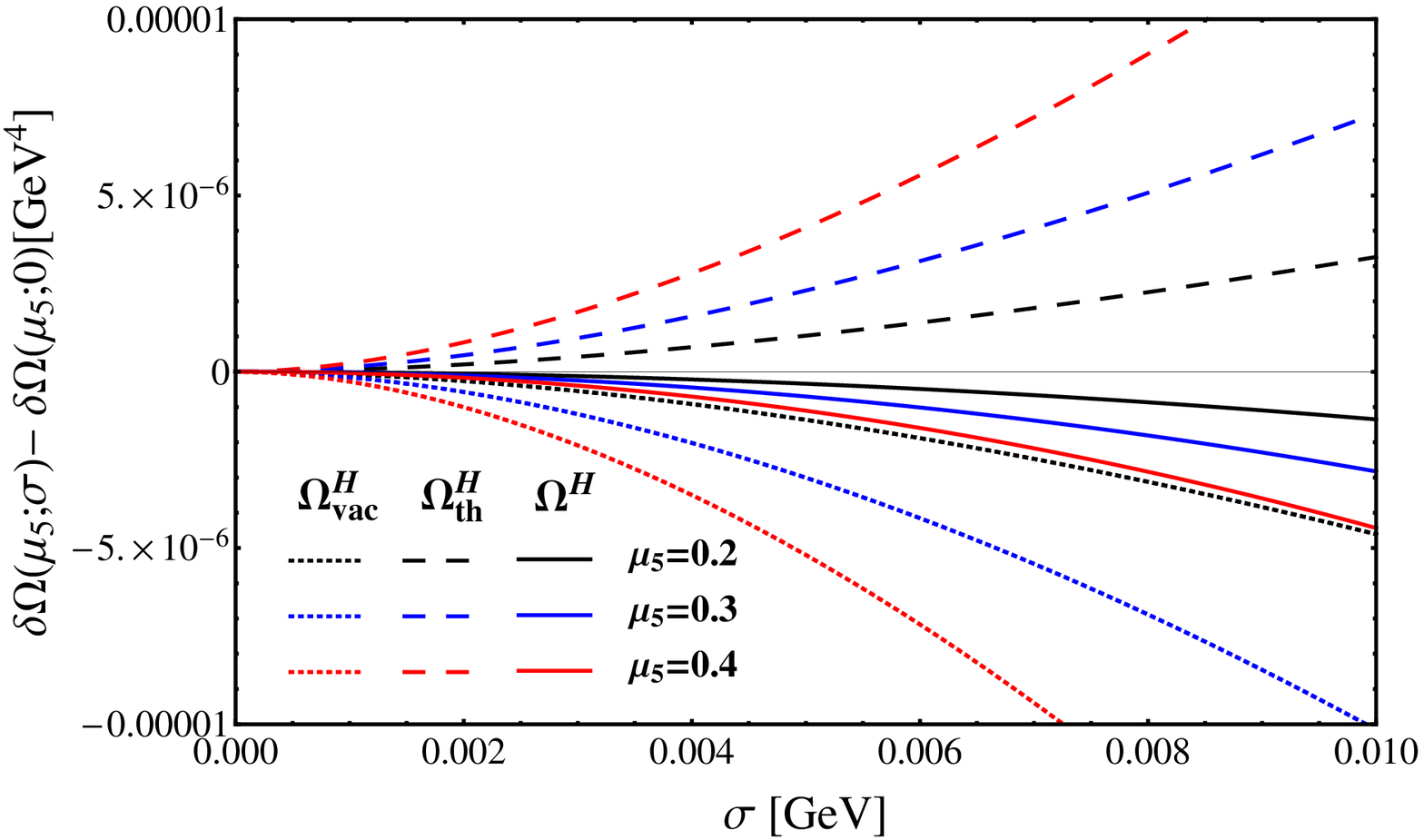}}
 \centerline{(a) $T=50$ MeV }
\vfill
 \centerline{\includegraphics[scale=0.49]{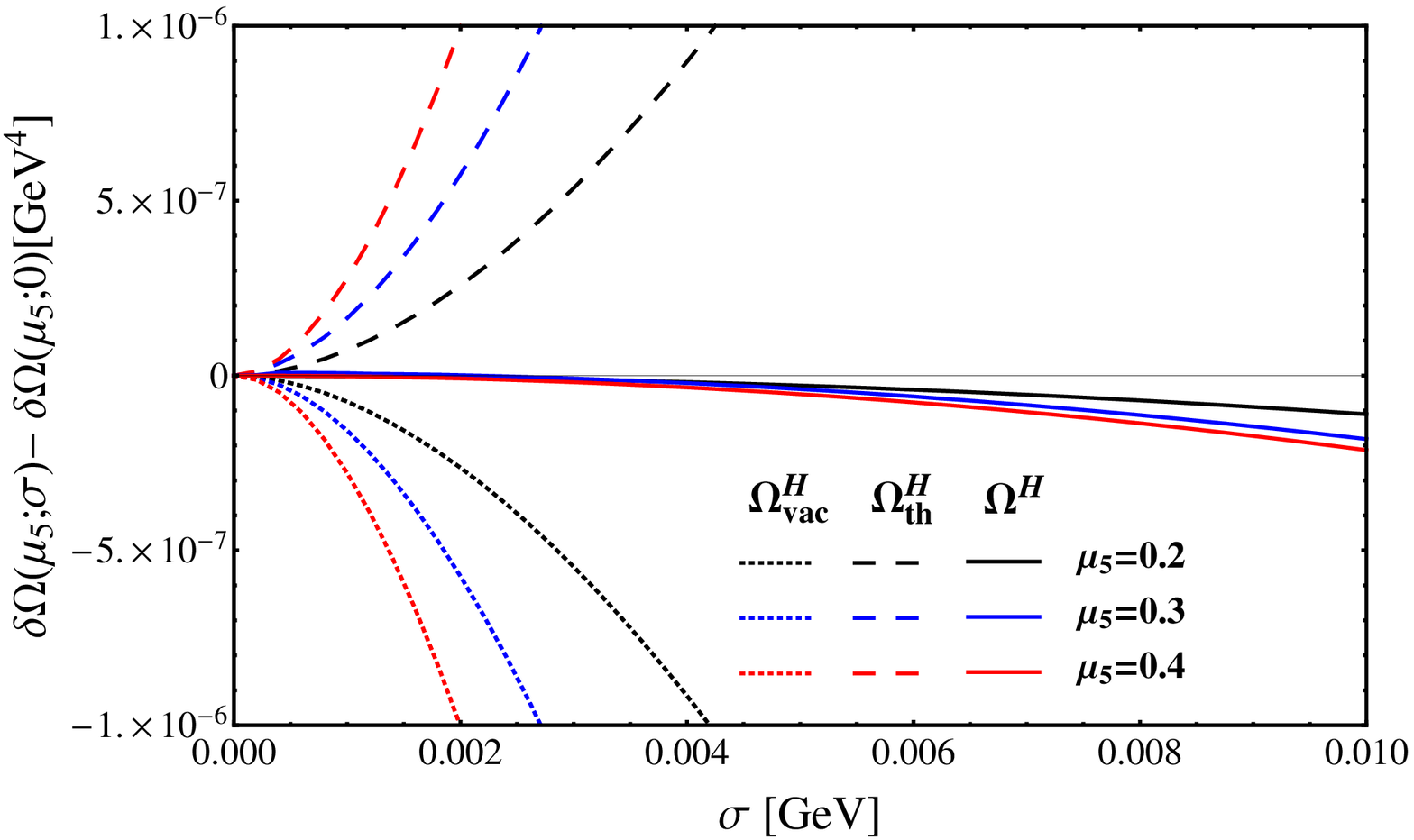}}
 \centerline{(b) $T=170$ MeV }

\caption{ $\delta\Omega^H_{vac}(\mu_5;\sigma)-\delta\Omega^H_{vac}(\mu_5;0)$, $\delta\Omega^H_{th}(\mu_5;\sigma)-\delta\Omega^H_{th}(\mu_5;0)$ and $\delta\Omega^H_{}(\mu_5;\sigma)-\delta\Omega^H_{}(\mu_5;0)$ as a function of $\sigma$ for $T=50$ and $170$ MeV at $\mu_5=0.2,~0.3~\text{and}~0.4$ GeV
near $\sigma=0$}
\label{fig:deltaO:T}
\end{figure}

And then, we have to resort to numerical calculations to investigate whether this result can be generalized for an arbitrary value of $\mu_5$ ($\mu_5 < \Lambda$ still) by using $\delta\Omega^H_{th}(\mu_5)$. In Fig.~\ref{fig:deltaO:T}, we plot the $\delta\Omega^H_{vac}(\mu_5;\sigma)-\delta\Omega^H_{vac}(\mu_5;0)$, $\delta\Omega^H_{th}(\mu_5;\sigma)-\delta\Omega^H_{th}(\mu_5;0)$ and $\delta\Omega^H_{}(\mu_5;\sigma)-\delta\Omega^H_{}(\mu_5;0)$ as a function of $\sigma$ for $T=50$ and $170$ MeV at $\mu_5=0.2,~0.3~\text{and}~0.4$ GeV
near $\sigma=0$. From Fig.~\ref{fig:deltaO:T}, for $T= 50 $ and $170$ MeV, one can find that the contribution of $\delta\Omega^H_{vac}$ to the coefficient of $\sigma^2$ term is negative, while that of $\delta\Omega^H_{th}$ is positive in contrast. It means that $\delta\Omega^H_{vac}$ helps to increase the chiral condensate, whereas $\delta\Omega^H_{th}$ helps to decrease the chiral condensate, based on previous discussions. And the sum of these two parts, i.e. $\delta\Omega^H=\delta\Omega^H_{vac}+\delta\Omega^H_{th}$, is negative and its magnitude increases with $\mu_5$. Hence, we find $\delta\Omega^H$ tends to enhance the chiral condensate at any temperatures and the enhancement increases with $\mu_5$ by more detailed numerical calculations, when $\mu_5<\mu_5^*$ ($0.4~ \text{GeV} <\mu_5^*<0.5~\text{GeV}$). And thus we can assume that $\delta\Omega^H$ contribute a term $-b(\mu_5)\sigma^2$, where $b(\mu_5)$ is a positive parameter increasing with $\mu_5$ and $b(\mu_5=0)=0$, to the thermodynamical potential. By using Eq.~(\ref{eq:H:highT}),
the coefficient of $\Omega^H(\mu_5)=\Omega^H(\mu_5=0)+\delta\Omega^H$ becomes
$(\frac{1}{4G_S}-\frac{3\Lambda^2}{2\pi^2}-b(\mu_5)^2+\chi^2(\Lambda)\frac{T^2}{2})$, so we can obtain the corresponding critical temperature $T^H_c(\mu_5)=\frac{1}{\chi(\Lambda)}\sqrt{(\frac{3\Lambda^2}{\pi^2}+b(\mu_5)- \frac{1}{2G_S})}>T^H_c(\mu_5=0)$ and the critical temperature increases with $\mu_5$ under this condition.

\begin{figure}
 \centerline{\includegraphics[scale=0.49]{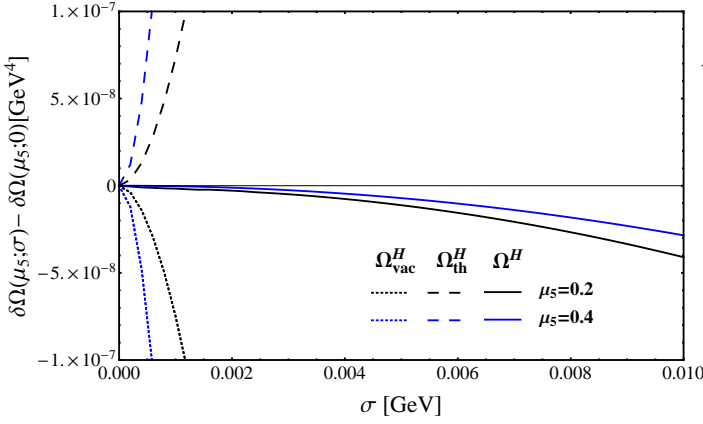}}

\caption{$\delta\Omega^H_{vac}(\mu_5;\sigma)-\delta\Omega^H_{vac}(\mu_5;0)$, $\delta\Omega^H_{th}(\mu_5;\sigma)-\delta\Omega^H_{th}(\mu_5;0)$ and $\delta\Omega^H_{}(\mu_5;\sigma)-\delta\Omega^H_{}(\mu_5;0)$ as a function of $\sigma$ for $T=200$ MeV at $\mu_5=0.2,~\text{and}~0.4$ near $\sigma=0$ }
\label{fig:deltaO:T200}
\end{figure}

Why the chiral chemical potential turns to reduce the critical temperature when $\mu_5>\mu_5^*$? We also could find the explanations in Fig.~\ref{fig:deltaO:T200}, by displaying the $\delta\Omega^H_{vac}(\mu_5;\sigma)-\delta\Omega^H_{vac}(\mu_5;0)$, $\delta\Omega^H_{th}(\mu_5;\sigma)-\delta\Omega^H_{th}(\mu_5;0)$ and $\delta\Omega^H_{}(\mu_5;\sigma)-\delta\Omega^H_{}(\mu_5;0)$ as a function of $\sigma$ for $T=200$ MeV at $\mu_5=0.2~\text{and}~0.4$ GeV near $\sigma=0$, respectively. Why we choose $T=200$ MeV? This is because the critical temperatures at $\mu_5=0.2$ and $0.4$ GeV are close to this value. Obviously, although the contributions of $\delta\Omega^H$ to the coefficient of $\sigma^2$ term for both $\mu_5=0.2$ and $0.4$ GeV are negative, the magnitude of the contribution at $\mu_5=0.4$ is smaller than that at $\mu_5=0.2$, which is different from the case at lower temperatures. It means that $T^H_c(\mu_5=0.2 \text{GeV})$ and  $T^H_c(\mu_5=0.4 \text{GeV})$ are larger than $T^H_c(\mu_5=0)$, but $T^H_c(\mu_5=0.2 \text{GeV})>T^H_c(\mu_5=0.4 \text{GeV})$, if we take similar discussions in the previous paragraph. And it is verified by the numerical simulations that $T^H_c(\mu_5=0.2 \text{GeV})=198.7$ MeV, while  $T^H_c(\mu_5=0.4 \text{GeV})=198.2$ MeV.

As for $\Omega^{H'(2)}_{th}$, although it decreases the chiral condensate at finite temperature also, it is not possible to determine whether its contribution is larger than that of $\Omega^{H'(2)}_{vac}$ around $T_c$ through small $\mu_5$ expansion because of its different upper limit of the integral as compared to $\Omega^{H(2)}_{th}$. But it is find that, by making use of the high-temperature expansion of the thermodynamical potential $\Omega^{H'}$ at finite $\mu_5$, we reveal the mechanism why $\mu_5$ helps to lower the critical temperature under the regularization scheme of $\Omega^{H'}$ analytically, which is given in the next subsection.

\subsubsection{High-temperature expansion with nonzero $\mu_5$}
\label{high:T:expansion}
Note that the results from $\Omega^{H'}$ are consistent with those from $\Omega^{S}$, so here we just show the reasons why $\Omega^{H'}$ is different from $\Omega^{H}$ at finite temperature through analytical derivation.

From Eq.~\ref{eq:Omega:H}, we could find that $\Omega^{H'}_{vac}$ takes the exactly same form as $\Omega^{H}_{vac}$:
\beqn
\label{eq:Omega:H:vacuum}
\Omega^{H/H'}_{vac}
&=&\frac{\sigma^2}{4G_S}-N_c N_f  \sum_{s=\pm}\int^{\Lambda}_{0}\frac{d p}{2\pi^2} \, p^2 \omega_{s}\,,\nonumber \\
&=&\frac{\sigma^2}{4G_S}-\frac{N_c N_f}{2\pi^2}\times\frac{1}{24}\Bigg\{\sqrt{(\Lambda+\mu_5)^2+\sigma^2}\Big[2(\Lambda\nonumber \\
&&+\mu_5)
(3\Lambda^2-2\Lambda\mu+\mu_5^2)+(3\Lambda-13\mu_5)\sigma^2\Big]\nonumber \\
&&+\sqrt{(\Lambda-\mu_5)^2+\sigma^2}\Big[2(\Lambda-\mu_5)
(3\Lambda^2+2\Lambda\mu+\mu_5^2)\nonumber \\
&&+(3\Lambda+13\mu_5)\sigma^2\Big]+6\sigma^2(\sigma^2-4\mu_5^2)\nonumber \\
&&\times
\ln\Big[\frac{\sigma}{2\sqrt{(\Lambda-\mu_5)(\Lambda+\mu_5)}}\Big]\Bigg\}.
\eeqn

When $T$ approaches $T_c$, if we have $\sigma \ll T$ and $\mu_5 \ll T$ (of course, $\sigma \ll \Lambda$ and $\mu_5 \ll \Lambda$ also), we can expand $\Omega^{H'}_{vac}$ ($\Omega^{H}_{vac}$) in powers of $\sigma/\Lambda$ to the second order:
\beqn
\Omega^{H'}_{vac}
&=&\frac{\sigma^2}{4G_S}-\frac{N_c N_f}{2\pi^2}\times\frac{1}{24}\Bigg\{\frac{6(\Lambda^3-\Lambda^2\mu_5
-4\Lambda\mu_5^2-2\mu_5^3)}{\Lambda+\mu_5}\nonumber \\
&&+
\frac{6(\Lambda^3+\Lambda^2\mu_5-4\Lambda\mu_5^2+2\mu_5^3)}{\Lambda-\mu_5}
-24\mu_5^2\nonumber \\
&&\times\ln\Big[\frac{\sigma}{2\sqrt{(\Lambda-\mu_5)(\Lambda+\mu_5)}}\Big]\Bigg\}\sigma^2+...
\eeqn
Thus, with considering $\mu_5 \ll \Lambda$, we get in the $\sigma\rightarrow 0$ limit:
\beqn
\label{eq:H':vac}
\delta\Omega^{H'}_{vac}(\mu_5;\sigma)-\delta\Omega^{H'}_{vac}(\mu_5;0)
&=&\frac{N_c N_f}{2\pi^2}\times\mu_5^2\bigg(\ln\frac{\sigma}{2\Lambda}+1\bigg)\sigma^2. \nonumber \\
\eeqn
Since $\big(\ln\frac{\sigma}{2\Lambda}+1\big)$ is negative near $T_c$, the contribution of $\delta\Omega^{H'}_{vac}$ to the coefficient of $\sigma^2$ term is negative near $\sigma=0$, which helps to increase the chiral condensate.

\begin{figure}
 \centerline{\includegraphics[scale=0.49]{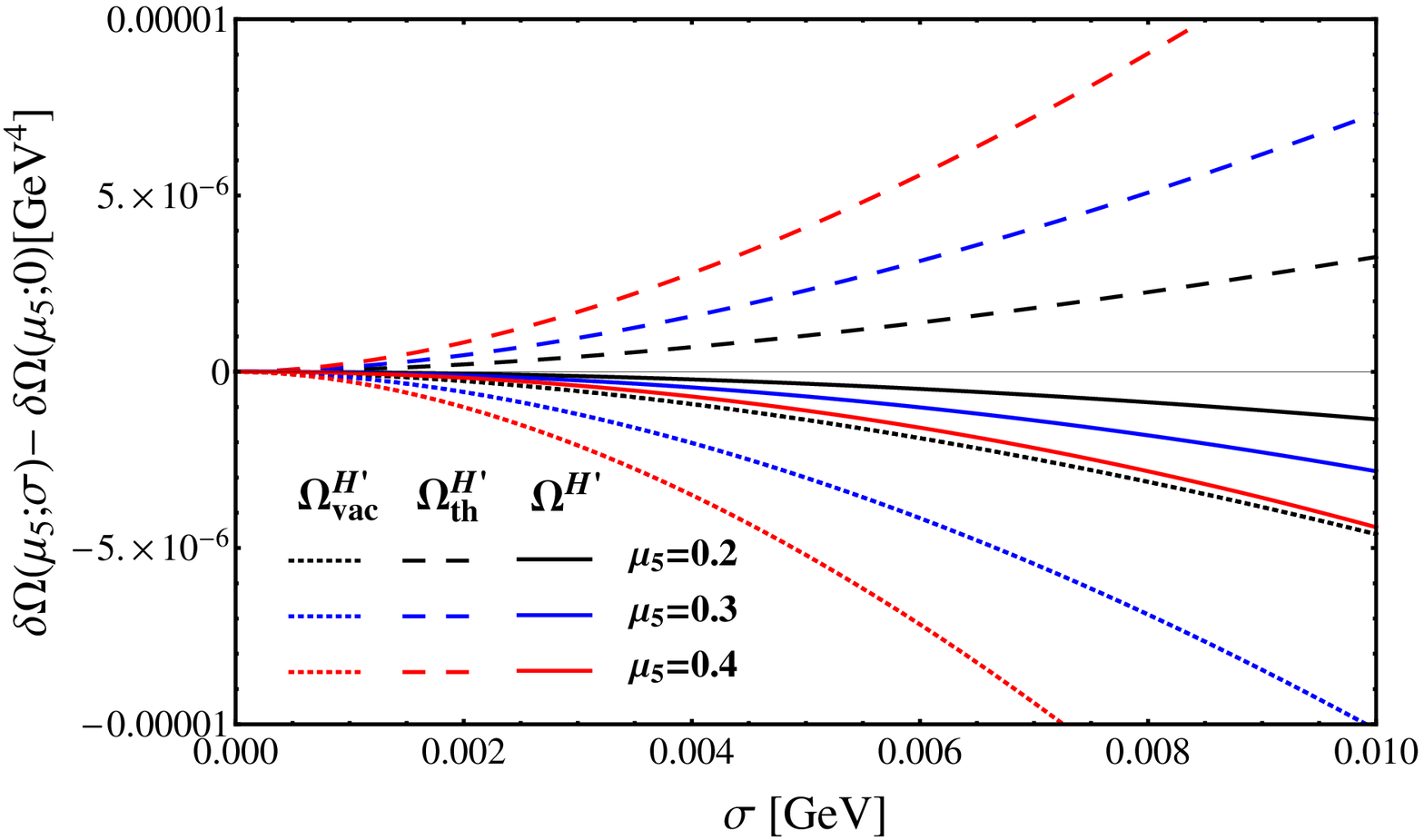}}
 \centerline{(a) $T=50$ MeV }
\vfill
 \centerline{\includegraphics[scale=0.49]{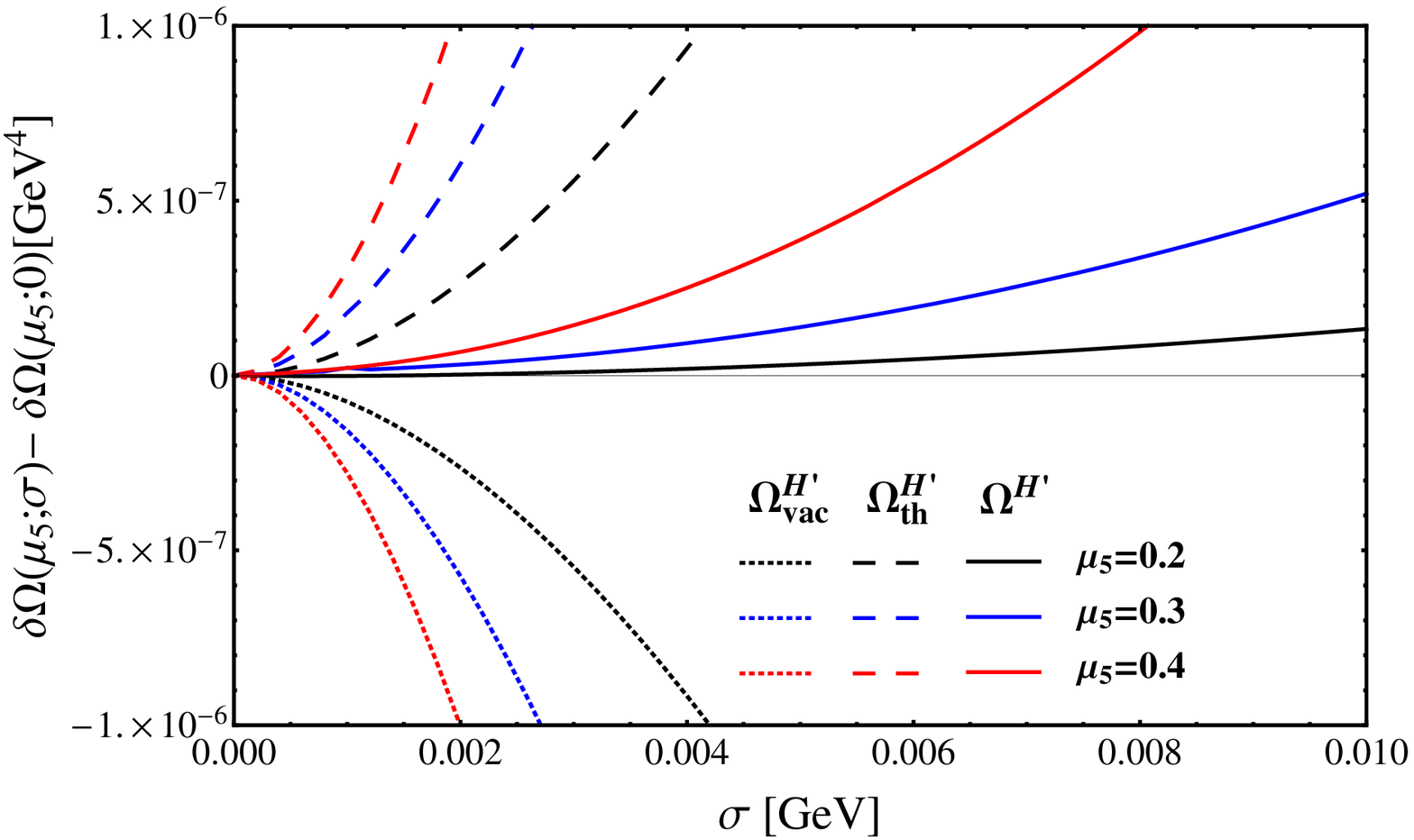}}
 \centerline{(b) $T=170$ MeV }

\caption{ $\delta\Omega^{H'}_{vac}(\mu_5;\sigma)-\delta\Omega^{H'}_{vac}(\mu_5;0)$, $\delta\Omega^{H'}_{th}(\mu_5;\sigma)-\delta\Omega^{H'}_{th}(\mu_5;0)$ and $\delta\Omega^{H'}_{}(\mu_5;\sigma)-\delta\Omega^{H'}_{}(\mu_5;0)$ as a function of $\sigma$ for $T=50$ and $170$ MeV at $\mu_5=0.2,~0.3~\text{and}~0.4$ GeV
near $\sigma=0$}
\label{fig:deltaOH':T}
\end{figure}

As for the thermal part, we have
\beqn
\Omega^{H'}_{th}
\label{eq:Omega:H':thermal}
&=&
-2N_c N_f\,T \sum_{s=\pm}\int^{\infty}_{0}\frac{d p}{2\pi^2}\, p^2\ln\bigl( 1+\, e^{-\beta \omega_{s}} \bigr)\,.\nonumber \\
\eeqn
When at high temperatures ($\sigma \ll T$ and $\mu_5 \ll T$), we can expand $\Omega^{H'}_{th}$ as follows (we use the same strategy in Refs.~\cite{Dolan:1973qd,Haber:1981fg,Haber:1981tr}):
\beqn
\Omega^{H'}_{th}
&=&
-2N_c N_f\Bigg\{(\frac{\mu_5^2}{12}-\frac{\sigma^2}{24})T^2+\frac{\sigma^4}{16\pi^2}\Big[\ln(\frac{\pi T}{\sigma})-\gamma_E+\frac{3}{4}\Big]\nonumber \\&&+\frac{1}{4\pi^2}\Big[\gamma_E -\frac{1}{2}-\ln(\frac{\pi T}{\sigma})\Big]\mu_5^2 \sigma^2+\frac{7\pi^2T^4}{360}\nonumber \\&&+\mathcal {O}(\frac{\sigma^6}{T^2},\frac{\sigma^4\mu_5^2}{T^2})\Bigg\},
\eeqn
Similar to $\Omega^{H'}_{vac}$, in the $\sigma\rightarrow 0$ limit, we obtain
\beqn
\label{eq:H':the}
\delta\Omega^{H'}_{th}(\mu_5;\sigma)-\delta\Omega^{H'}_{th}(\mu_5;0)
&=&
\frac{N_c N_f}{2\pi^2}\Big[\ln(\frac{\pi T}{\sigma})-\gamma_E\nonumber \\&&+\frac{1}{2}\Big]\mu_5^2 \sigma^2.
\eeqn
Since $T\gg \sigma$ near $T_c$, the contribution of the thermal part to the coefficient of $\sigma^2$ term is positive near $\sigma=0$ and it helps to decrease the chiral condensate.

And then, combining Eq.~(\ref{eq:H':the}) with Eq.~(\ref{eq:H':vac}), we can find that
\beqn
\label{eq:H':tot}
\delta\Omega^{H'}(\mu_5;\sigma)-\delta\Omega^{H'}(\mu_5;0)
&=&
\frac{N_c N_f}{2\pi^2}\Big[\ln(\frac{\pi T}{2\Lambda})-\gamma_E\nonumber \\&&+\frac{3}{2}\Big]\mu_5^2 \sigma^2.
\eeqn
When $T$ is large enough, its coefficient of $\sigma^2$ term will become positive from negative. It indicates that, when at low temperatures, $\delta\Omega^{H'}$ helps to increase the chiral condensate, whereas when at enough high temperatures, this will make a change and $\delta\Omega^{H'}$ helps to decrease the chiral condensate. Moreover, from Eq.~(\ref{eq:H':tot}), the magnitude of the coefficient will increase with $\mu_5$ when at enough high temperatures and it means the critical temperature will decrease with $\mu_5$ by taking the analysis similar to $\delta\Omega^{H}$. Assuming $T_c=T_c^0+\delta T$ at small $\mu_5$, where
$T_c^0=T_c(\mu_5=0)$, we can find $\delta T$ satisfies
\beqn
(\frac{1}{4G_S}-\frac{3\Lambda^2}{2\pi^2}+\frac{T_c^2}{2})+\frac{N_c N_f}{2\pi^2}\Big[\ln(\frac{\pi T_c}{2\Lambda})-\gamma_E+\frac{3}{2}\Big]\mu_5^2=0,\nonumber \\&&
\eeqn
and then we have $\delta T=-\frac{3\mu_5^2}{\pi^2(T_c^0)^2}\Big[\ln(\frac{\pi T_c^0}{2\Lambda})-\gamma_E+\frac{3}{2}\Big]T_c^0<0$, since it is to check
$\ln(\frac{\pi T_c^0}{2\Lambda})-\gamma_E+\frac{3}{2}>0$. This result was also obtained in Ref.~\cite{Gatto:2011wc}.

In order to investigate the validity of our analysis at relatively large $\mu_5$ values, we resort to numerical computations to plot $\delta\Omega^{H'}_{vac}(\mu_5;\sigma)-\delta\Omega^{H'}_{vac}(\mu_5;0)$, $\delta\Omega^{H'}_{th}(\mu_5;\sigma)-\delta\Omega^{H'}_{th}(\mu_5;0)$ and $\delta\Omega^{H'}_{}(\mu_5;\sigma)-\delta\Omega^{H'}_{}(\mu_5;0)$ as a function of $\sigma$ for $T=50$ and $170$ MeV at $\mu_5=0.2,~0.3~\text{and}~0.4$ GeV
near $\sigma=0$ in Fig.~\ref{fig:deltaOH':T}. From Fig.~\ref{fig:deltaOH':T},
we notice that, when $T=50$ MeV (a low temperature), the sum of the vacuum and thermal parts, $\delta\Omega^{H}$, is negative and the magnitude of the coefficient of the $\sigma^2$ term increase with $\mu_5$. Hence, we can conclude that the chiral condensate make an enhancement as $\mu_5$ increases. When at $T=170$ MeV (close to $T^{H'}_c (\mu_5=0)$), it is obvious that the contribution of $\delta\Omega^{H}$ to the coefficient of the $\sigma^2$ becomes positive and its magnitude increases with $\mu_5$ also. Consequently, the critical temperature will decrease with $\mu_5$.

\section{conclusions}
\label{conclusion}

In this paper, we have studied the effects of the chiral chemical potential $\mu_5$ on the chiral phase transition for
three different regularization schemes in the NJL model. Based on the analytical and numerical analysis of the
thermodynamical potential $\Omega$, we find that the coefficient of $\sigma^2$ term for the potential $\Omega$ reflects
the behavior of the chiral condensate and the critical temperature. Thus, we compare the modifications of the coefficients
of $\sigma^2$ terms made by $\mu_5$, for the thermodynamical potentials in three regularization schemes,
i.e. $\Omega^{S}$, $\Omega^{H}$ and $\Omega^{H'}$. In this way, we have found the reason for the puzzle that why the same
NJL model in Refs.~\cite{Fukushima:2010fe,Gatto:2011wc} and \cite{Cao:2015}, due to the choice of the regularization scheme,
gave two opposite results about the dependence
of the critical temperature on $\mu_5$ totally distinct from each other. This is because the conventional hard-cutoff
regularization scheme includes a hard cutoff function $\theta(\Lambda-|\bm p|)$ for the thermal part of the thermodynamical
potential, so that its thermal part $\Omega^{H}_{th}$
does not contain all the contribution from the chiral chemical potential $\mu_5$. In fact, the thermal part of the
thermodynamical potential is convergent, so there is no need to introduce any cutoff function for it at all. Therefore, the results
obtained from the soft-cutoff and revised hard-cutoff regularization schemes are more reasonable.

In our study, we have first investigated the effect of the chiral chemical potential on the chiral condensate for the vacuum part and the
thermal part of the thermodynamical potential, respectively. Our results show a competition between these two parts that, the vacuum part
tends to increase the chiral condensate, while the thermal part tends to decrease the chiral condensate. Relying on both analytical
analysis of the small $\mu_5$ expansion and
numerical computations, it is found that at zero and low temperatures, if $mu_5<mu_5^{thr}$, the contribution of the vacuum part is greater than
that of the thermal part for all three regularization schemes, so that the chiral chemical potential catalyze the chiral symmetry breaking,
which is consistent with the result of all previous Refs.~\cite{Fukushima:2010fe,Chernodub:2011fr,Gatto:2011wc,Cao:2015,
Wang:2015tia,Xu:2015vna,Braguta:2014ira}. When at high temperatures
around $T_c$, different regularization schemes give two different results. For $\Omega^{H}$, perturbative expansion in the small $\mu_5$ limit, as well as numerical
calculations, explicitly tells us the contribution of the thermal term is always less than that of the vacuum term at any temperatures because of its regularization scheme,
which is also verified by the numerical results for $\mu_5<\mu_5^*$. As for $\Omega^{S}$ and $\Omega^{H'}$, the numerical calculations show us that
the contribution of the thermal term will exceed that of the vacuum term when the temperature is high enough. Moreover, we should mention that, even
at zero temperature, the dependence of the chiral condensate as a function of $\mu_5$ is not monotonically increasing, which is different from all previous results. When $\mu_5>\mu_5^*$, the quark condensate turns to
decrease with $\mu_5$, although its value is still larger than that in the vacuum.

We have also computed the effect of the chiral chemical potential on the chiral critical temperature. On the basis of the discussions about
the corrections of $\mu_5$ on the coefficient of $\sigma^2$ term for the potential $\Omega$, we find that, for $\Omega^{H}$, the critical temperature
$T_c$ increases with $\mu_5$ when $\mu_5<\mu_5^*$, while $T_c$ decreases with $\mu_5$ when $\mu_5>\mu_5^*$. The latter is not in agreement
with the result in Ref.~\cite{Cao:2015}, and this is because that in Ref.~\cite{Cao:2015}, its maximum value of $\mu_5$ is just $0.5\Lambda$ and
less than $\mu_5^*$. Depending on the similar analysis,
we obtain that the critical temperature is always decreasing function of $\mu_5$ for $\Omega^{S}$ and $\Omega^{H'}$. Therefore, for the conventional
hard-cutoff regularization scheme, its increasing behavior of the critical temperature as a function of the chiral chemical potential is caused by
the incomplete contribution of the thermal part.

In the future, it is important to extend our analysis to the $2+1$ flavors and take into account the effect of
finite current quark mass. On the other hand,
it is of interest to explore the underling reasons for the order of the chiral phase transition with
finite chiral chemical potential in the NJL model with different regularization schemes. Moreover,
why the behavior of the critical temperature in both the NJL model~\cite{Fukushima:2010fe,Gatto:2011wc} and some other effective models~\cite{Chernodub:2011fr} is different from
that in DSEs~\cite{Xu:2015vna} and lattice QCD~\cite{Braguta:2014ira} is still an open question.
How to choose a proper regularization scheme may play an important role.

\vskip 0.2cm
\section*{Acknowledgement}

We thank valuable discussions with M. Chernodub and T. Kojo.
This work is supported by the NSFC under Grant No. 11275213, and 11261130311(CRC 110 by DFG and NSFC),
CAS key project under Grant No. KJCX2-EW-N01, and the Youth Innovation Promotion Association of CAS.
L.Yu is partially supported by China Postdoctoral Science Foundation under
Grant No. 2014M550841 and the Seeds Funding of Jilin University.

\end{document}